\documentclass[preprint,3p,times,onecolumn, sort&compress]{elsarticle}
\pdfoutput=1
\usepackage{etoolbox}
\usepackage[utf8]{inputenc}
\usepackage[colorlinks]{hyperref}
\usepackage{graphicx}
\usepackage{tabularx}
\usepackage{lipsum}
\usepackage{wrapfig}
\usepackage{amsmath}
\usepackage{float}
\usepackage{xfrac}
\usepackage{subfigure}
\usepackage{gensymb}
\usepackage{hyperref}
\usepackage{cleveref}
\usepackage{chngcntr}
\usepackage{multirow}
\usepackage{makecell}
\usepackage{array}
\usepackage{booktabs}
\usepackage[none]{hyphenat} 
\usepackage{enumitem}
\usepackage{ragged2e}
\usepackage{amssymb}
\usepackage[section]{placeins}
\usepackage{siunitx}
\usepackage{epigraph}
\usepackage{lineno}
\usepackage{braket}
\usepackage{xparse,nameref}
\usepackage[usestackEOL]{stackengine}
\usepackage[svgnames]{xcolor}
\usepackage{mathtools}
\usepackage[version=4]{mhchem}
\usepackage{multirow}
\usepackage{makecell}
\usepackage{array}
\usepackage{booktabs}
\usepackage{textcase}
\usepackage[toc,page]{appendix}
\usepackage{pdfpages}
\usepackage{titlesec}
\usepackage{natbib}
\usepackage{tikz}
\usepackage{pgf}
\usepackage{pgfplots}
\usepackage{pgfplotstable}
\pgfplotsset{compat=newest}
\usetikzlibrary{calc, math, patterns, external, trees, positioning}
\pgfplotsset{select coords between index/.style 2 args={
    x filter/.code={
        \ifnum\coordindex<#1\fi
        \ifnum\coordindex>#2\fi
    }
}}

\titleformat{\subsection} {\normalfont\bfseries}{\thesubsection}{1em}{}
\titleformat{\paragraph} {\normalfont\bfseries\itshape}{\paragraph}{1em}{}
\usepackage{dcolumn}   

\usepackage[tablename=TABLE, labelsep=period, figurename=Figure, skip=0pt, font=small]{caption}

\newcommand{\Tc}{\ensuremath{\mathrm{T}_\text{c}}}
\newcommand{\ssm}{\ensuremath{^\text{SSM}}}
\newcommand{\FluxO}{\ensuremath{\Phi_{\ce{O}}}}

\newcommand{\FluxB}{\ensuremath{\Phi_{\ce{B}}}}

\makeatletter
\def\ps@pprintTitle{%
 \let\@oddhead\@empty
 \let\@evenhead\@empty
 \def\@oddfoot{}%
 \let\@evenfoot\@oddfoot}
\makeatother

\begin{document}
  \begin{frontmatter}
       \title{\textbf{Improved measurement of solar neutrinos from the Carbon-Nitrogen-Oxygen cycle by Borexino and its implications for the Standard Solar Model}}

\author[Munchen]{S.~Appel}
\author[Juelich]{Z.~Bagdasarian\fnref{Berkeley}}
\author[Milano]{D.~Basilico}
\author[Milano]{G.~Bellini}
\author[PrincetonChemEng]{J.~Benziger}
\author[LNGS]{R.~Biondi}
\author[Milano]{B.~Caccianiga}
\author[Princeton]{F.~Calaprice}
\author[Genova]{A.~Caminata}
\author[Virginia]{P.~Cavalcante\fnref{LNGSG}}
\author[Lomonosov]{A.~Chepurnov}
\author[Milano]{D.~D'Angelo}
\author[Peters]{A.~Derbin}
\author[LNGS]{A.~Di Giacinto}
\author[LNGS]{V.~Di Marcello}
\author[Princeton]{X.F.~Ding}
\author[Princeton]{A.~Di Ludovico\fnref{LNGSG}} 
\author[Genova]{L.~Di Noto}
\author[Peters]{I.~Drachnev}
\author[APC]{D.~Franco}
\author[Princeton,GSSI]{C.~Galbiati}
\author[LNGS]{C.~Ghiano}
\author[Milano]{M.~Giammarchi}
\author[Princeton]{A.~Goretti\fnref{LNGSG}}
\author[Juelich,RWTH]{A.S.~G\"ottel}
\author[Lomonosov,Dubna]{M.~Gromov}
\author[Mainz]{D.~Guffanti\fnref{Bicocca}}
\author[LNGS]{Aldo~Ianni}
\author[Princeton]{Andrea~Ianni}
\author[Krakow]{A.~Jany}
\author[Kiev]{V.~Kobychev}
\author[London,Atomki]{G.~Korga}
\author[Juelich,RWTH]{S.~Kumaran}
\author[LNGS]{M.~Laubenstein}
\author[Kurchatov,Kurchatovb]{E.~Litvinovich}
\author[Milano]{P.~Lombardi}
\author[Peters]{I.~Lomskaya}
\author[Juelich,RWTH]{L.~Ludhova}
\author[Kurchatov]{G.~Lukyanchenko}
\author[Kurchatov,Kurchatovb]{I.~Machulin}
\author[Mainz]{J.~Martyn}
\author[Milano]{E.~Meroni}
\author[Milano]{L.~Miramonti}
\author[Krakow]{M.~Misiaszek}
\author[Peters]{V.~Muratova}
\author[Kurchatov,Kurchatovb]{R.~Nugmanov}
\author[Munchen]{L.~Oberauer}
\author[Mainz]{V.~Orekhov}
\author[Perugia]{F.~Ortica}
\author[Genova]{M.~Pallavicini}
\author[Munchen]{L.~Papp}
\author[Juelich,RWTH]{L.~Pelicci}
\author[Juelich]{\"O.~Penek}
\author[Princeton]{L.~Pietrofaccia\fnref{LNGSG}}
\author[Peters]{N.~Pilipenko}
\author[UMass]{A.~Pocar}
\author[Kurchatov]{G.~Raikov}
\author[LNGS]{M.T.~Ranalli}
\author[Milano]{G.~Ranucci}
\author[LNGS]{A.~Razeto}
\author[Milano]{A.~Re}
\author[Juelich,RWTH]{M.~Redchuk\fnref{Padova}}
\author[LNGS]{N.~Rossi}
\author[Munchen]{S.~Sch\"onert}
\author[Peters]{D.~Semenov}
\author[Juelich]{G.~Settanta\fnref{ISPRA}}
\author[Kurchatov,Kurchatovb]{M.~Skorokhvatov}
\author[Juelich,RWTH]{A.~Singhal}
\author[Dubna]{O.~Smirnov}
\author[Dubna]{A.~Sotnikov}
\author[LNGS]{R.~Tartaglia}
\author[Genova]{G.~Testera}
\author[Peters]{E.~Unzhakov}
\author[LNGS,Aquila]{F.L.~Villante}
\author[Dubna]{A.~Vishneva}
\author[Virginia]{R.B.~Vogelaar}
\author[Munchen]{F.~von~Feilitzsch}
\author[Krakow]{M.~Wojcik}
\author[Mainz]{M.~Wurm}
\author[Genova]{S.~Zavatarelli}
\author[Dresda]{K.~Zuber}
\author[Krakow]{G.~Zuzel}

\fntext[Berkeley]{Present address: University of California, Berkeley, Department of Physics, CA 94720, Berkeley, USA}
\fntext[LNGSG]{Present address: INFN Laboratori Nazionali del Gran Sasso, 67010 Assergi (AQ), Italy}
\fntext[Padova]{Present address: Dipartimento di Fisica e Astronomia, Universit\`a degli Studi e INFN, Padova, Italy}
\fntext[ISPRA]{Present address: Istituto Superiore per la Protezione e la Ricerca Ambientale, 00144 Roma, Italy}
\fntext[Bicocca]{Present address: Dipartimento di Fisica, Universit\`a degli Studi e INFN Milano-Bicocca, 20126 Milano, Italy}
\address{\bf{The Borexino Collaboration}}

\address[APC]{AstroParticule et Cosmologie, Universit\'e Paris Diderot, CNRS/IN2P3, CEA/IRFU, Observatoire de Paris, Sorbonne Paris Cit\'e, 75205 Paris Cedex 13, France}
\address[Dubna]{Joint Institute for Nuclear Research, 141980 Dubna, Russia}
\address[Genova]{Dipartimento di Fisica, Universit\`a degli Studi e INFN, 16146 Genova, Italy}
\address[Krakow]{M.~Smoluchowski Institute of Physics, Jagiellonian University, 30348 Krakow, Poland}
\address[Kiev]{Kiev Institute for Nuclear Research, 03680 Kiev, Ukraine}
\address[Kurchatov]{National Research Centre Kurchatov Institute, 123182 Moscow, Russia}
\address[Kurchatovb]{ National Research Nuclear University MEPhI (Moscow Engineering Physics Institute), 115409 Moscow, Russia}
\address[LNGS]{INFN Laboratori Nazionali del Gran Sasso, 67100 Assergi (AQ), Italy}
\address[Milano]{Dipartimento di Fisica, Universit\`a degli Studi e INFN, 20133 Milano, Italy}
\address[Perugia]{Dipartimento di Chimica, Biologia e Biotecnologie, Universit\`a degli Studi e INFN, 06123 Perugia, Italy}
\address[Peters]{St. Petersburg Nuclear Physics Institute NRC Kurchatov Institute, 188350 Gatchina, Russia}
\address[Princeton]{Physics Department, Princeton University, Princeton, NJ 08544, USA}
\address[PrincetonChemEng]{Chemical Engineering Department, Princeton University, Princeton, NJ 08544, USA}
\address[UMass]{Amherst Center for Fundamental Interactions and Physics Department, University of Massachusetts, Amherst, MA 01003, USA}
\address[Virginia]{Physics Department, Virginia Polytechnic Institute and State University, Blacksburg, VA 24061, USA}
\address[Munchen]{Physik-Department, Technische Universit\"at  M\"unchen, 85748 Garching, Germany}
\address[Lomonosov]{Lomonosov Moscow State University Skobeltsyn Institute of Nuclear Physics, 119234 Moscow, Russia}
\address[GSSI]{Gran Sasso Science Institute, 67100 L'Aquila, Italy}
\address[Dresda]{Department of Physics, Technische Universit\"at Dresden, 01062 Dresden, Germany}
\address[Mainz]{Institute of Physics and Excellence Cluster PRISMA+, Johannes Gutenberg-Universit\"at Mainz, 55099 Mainz, Germany}
\address[Juelich]{Institut f\"ur Kernphysik, Forschungszentrum J\"ulich, 52425 J\"ulich, Germany}
\address[RWTH]{III. Physikalisches Institut B, RWTH Aachen University, 52062 Aachen, Germany}
\address[London]{Department of Physics, Royal Holloway University of London, Egham, Surrey,TW20 0EX, UK}
\address[Atomki]{Institute of Nuclear Research (Atomki), Debrecen, Hungary}
\address[Aquila]{Dipartimento di Scienze Fisiche e Chimiche, Universit\`a dell'Aquila, 67100 L'Aquila, Italy}

\begin{abstract}

We present an improved measurement of the CNO solar neutrino interaction rate at Earth obtained with the complete Borexino Phase-III dataset. The measured rate R$_{\rm CNO}$\,=\,$6.7^{+2.0}_{-0.8}$\,counts/(day$\,\cdot$\,100\,tonnes), allows us to exclude the absence of the CNO signal with about 7$\sigma$ C.L. The correspondent CNO neutrino flux is $6.6^{+2.0}_{-0.9} \times 10^8$\,cm$^{-2}$\,s$^{-1}$, taking into account the neutrino flavor conversion.
We use the new CNO measurement to evaluate the C and N abundances in the Sun with respect to the H abundance for the first time with solar neutrinos.
 Our result of $N_{\rm CN}$\,=\,$(5.78^{+1.86}_{-1.00})\times10^{-4}$ displays a $\sim$2$\sigma$ tension with the ``low metallicity" spectroscopic photospheric measurements. On the other hand, our result used together with the $^7$Be and $^8$B solar neutrino fluxes, also measured by Borexino, permits to disfavour at 3.1$\sigma$\,C.L. the ``low metallicity" SSM B16-AGSS09met as an alternative to the ``high metallicity" SSM B16-GS98.

\end{abstract}
     
\end{frontmatter}
\twocolumn 

{\it Introduction} --- The Sun is powered in its core by nuclear reactions converting hydrogen into helium. This fusion proceeds via two sequences, the proton-proton ($pp$) chain producing about 99\% of energy and the subdominant CNO cycle. Neutrinos ($\nu$'s), emitted in both sequences, escape the solar matter almost unperturbed, delivering to us a real-time picture of the solar core. Over the last 50 years, the experimental effort has succeeded to map all the reactions producing solar $\nu$'s in the $pp$ chain ($pp$, $pep$, $^7$Be, and $^8$B $\nu$'s, with the exception of the extremely small expected flux of $hep$ $\nu$'s)\,\cite{Cleveland:1998nv, SAGE:2009eeu, Kaether:2010ag, SK2006, SK2016, SNO:I, SNO:II, ppchain-nature} and recently to provide the first direct evidence of CNO $\nu$'s\,\cite{CNO-nature}. These results have been crucial for solar physics, providing a precise test of the Standard Solar Model (SSM, latest available SSM B16\,\cite{Vinyoles:2016djt}), as well as for particle physics, contributing to the discovery of the neutrino flavour conversion~\cite{SNO:I,SNO:II} and measurement of the oscillation parameters~\cite{Esteban:2020cvm}. Furthermore, since the CNO cycle is predicted to be the dominant stellar hydrogen burning mechanism in the universe\,\cite{Salaris}, its detection sets a milestone for experimental astrophysics.

The CNO cycle consists of two sub-cycles, called CN and NO: at the relatively low temperature of the solar core, sub-cycle CN is largely dominant at $\sim$99\% level and produces neutrinos from the $\beta$-decays of $^{15}$O and $^{13}$N. In the CNO cycle, the fusion is catalyzed by carbon (C), nitrogen (N), and oxygen (O) and thus provides direct information on the metallicity of the Sun's core, i.e., its abundance of elements heavier than helium.

Metallicity is a key input of the SSMs and is determined experimentally by the spectral analysis of the photosphere, sometimes complemented by studies of meteorites: while measurements from the past two decades (AGSS09met\,\cite{AGSS09,SMaccretion}, C11\,\cite{C11}, AAG21\,\cite{AAG21}) have been suggesting a lower content of heavy elements with respect to the earlier ones (GS98\,\cite{GS98}), the most recent MB22\,\cite{MB22} results point to a higher value.
Noticeably, SSMs implementing the class of ``low-metallicity" compositions
fail to reproduce helioseismological measurements, while ``high-metallicity'' ones
are in better agreement with them\,\cite{Vinyoles:2016djt,MB22}.

Metallicity impacts the SSM predictions of $^8$B, $^7$Be, and CNO $\nu$ fluxes significantly, but in an indirect way. The  metal content affects the solar opacity, which in turn  impacts the Sun's temperature profile, which ultimately controls the rate of nuclear reactions and thus $\nu$ emission. Thus deriving information on metallicity from the measurements of solar $\nu$'s presents a certain degree of ambiguity.
However, in this respect, the CN cycle which is catalyzed by the C and N, is special: its flux has an additional, almost linear dependence on the abundances of these metals in the solar core, providing a unique handle for their non-ambiguous determination.

\begin{figure}[t]
    \centering
    \includegraphics[width=0.49\textwidth]{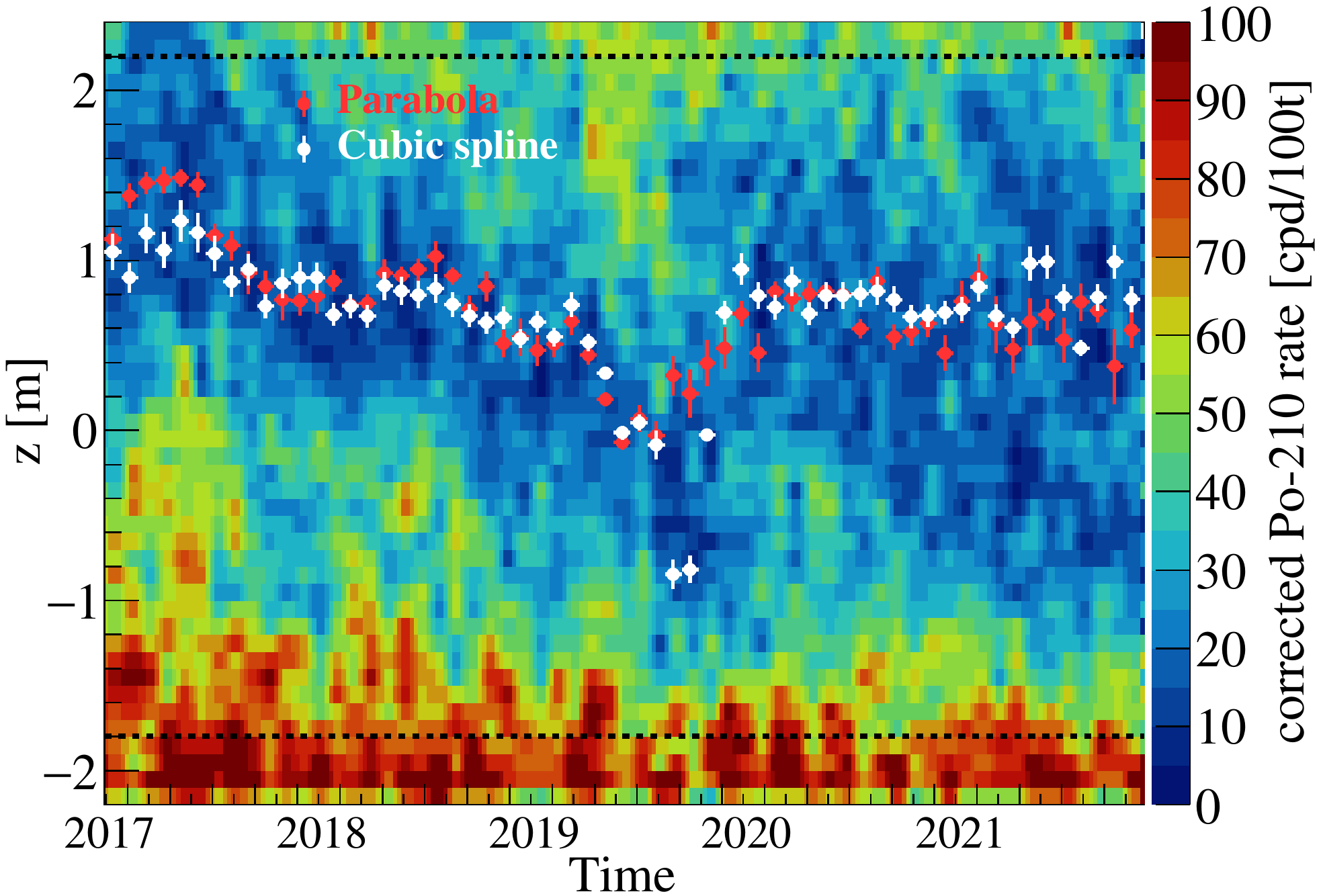}\vspace{2mm}
    \caption{Time evolution of the $^{210}$Po rate in the detector, visualized in terms of cylindrical $z$-slices of 0.1\,m height and radius $\rho^2 = (x^2 + y^2) <$\,2\,m$^{2}$. The horizontal black dashed lines represent the $z$-cut used in the CNO analysis. The Low Polonium Field centers obtained from the monthly paraboloid fits with (white) and without (red) a cubic spline along the $z$-axis are also shown.}   
    \label{fig:210Po_map}
\end{figure}

In this letter we present an improved measurement of the CNO $\nu$ interaction rate, obtained with the complete Borexino Phase-III dataset and a significantly increased precision, when compared to~\cite{CNO-nature}. We include this new result in the global analysis of all solar neutrino and KamLAND reactor antineutrino data. We compare the resulting solar neutrino fluxes to the predictions of SSM B16, using either GS98 or AGSS09met metallicity\,\cite{Vinyoles:2016djt} as an input.
Finally, we combine the CNO measurement with the $^8$B flux obtained from the global analysis to determine the C and N abundance directly. As discussed below, this procedure has an advantage of exploiting the precise measurement of $^8$B neutrino flux as a solar thermometer, minimizing  the uncertainties due to the metallicity/opacity degeneracy, and provides an estimation of metallicity which is independent from the spectroscopic data for the first time.

\begin{figure*}[t!]
    \subfigure[]
       {\includegraphics[width=0.5\textwidth]{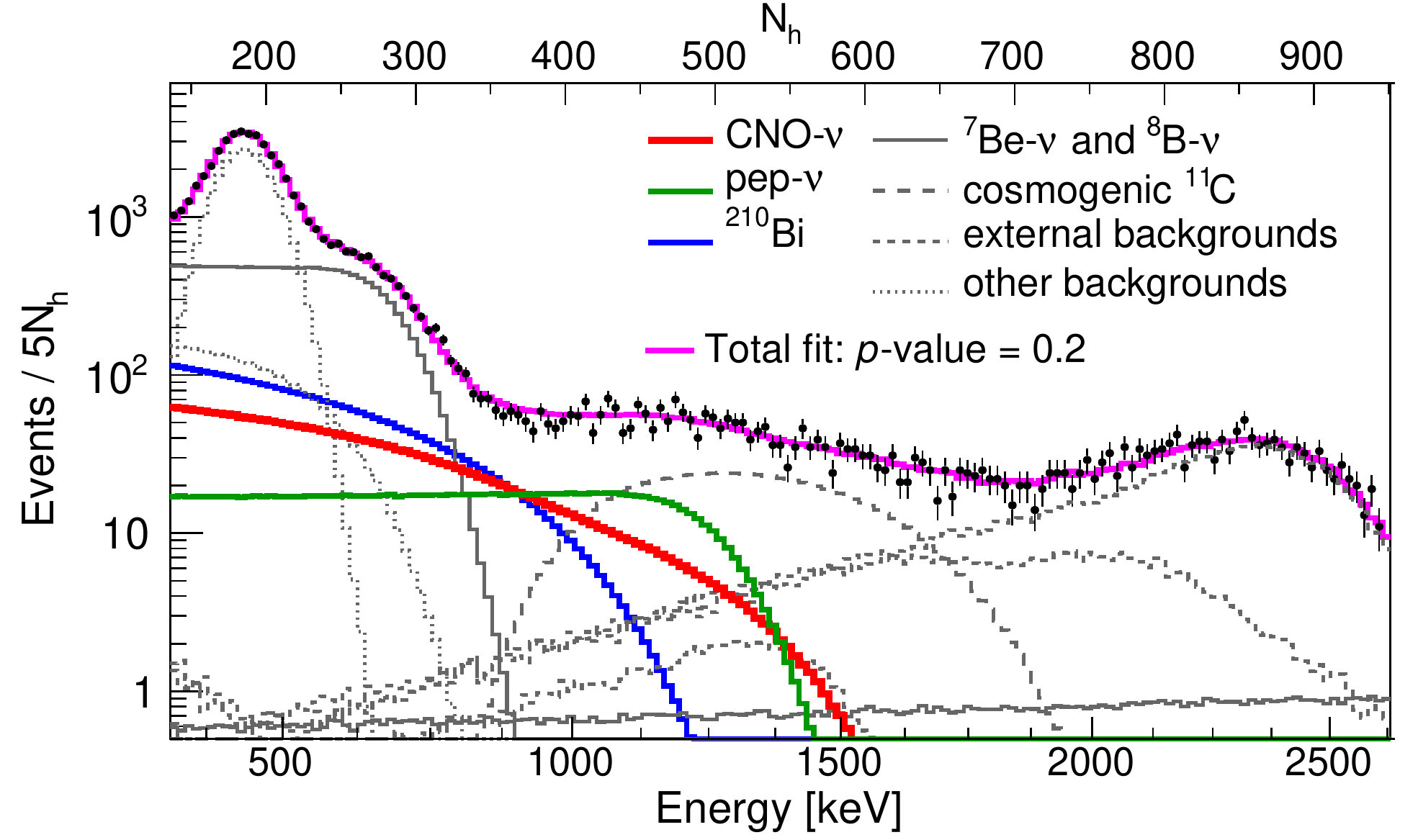}
    \label{fig:TFCsubtracted}}
    \subfigure[] {\includegraphics[width=0.45\textwidth, ]{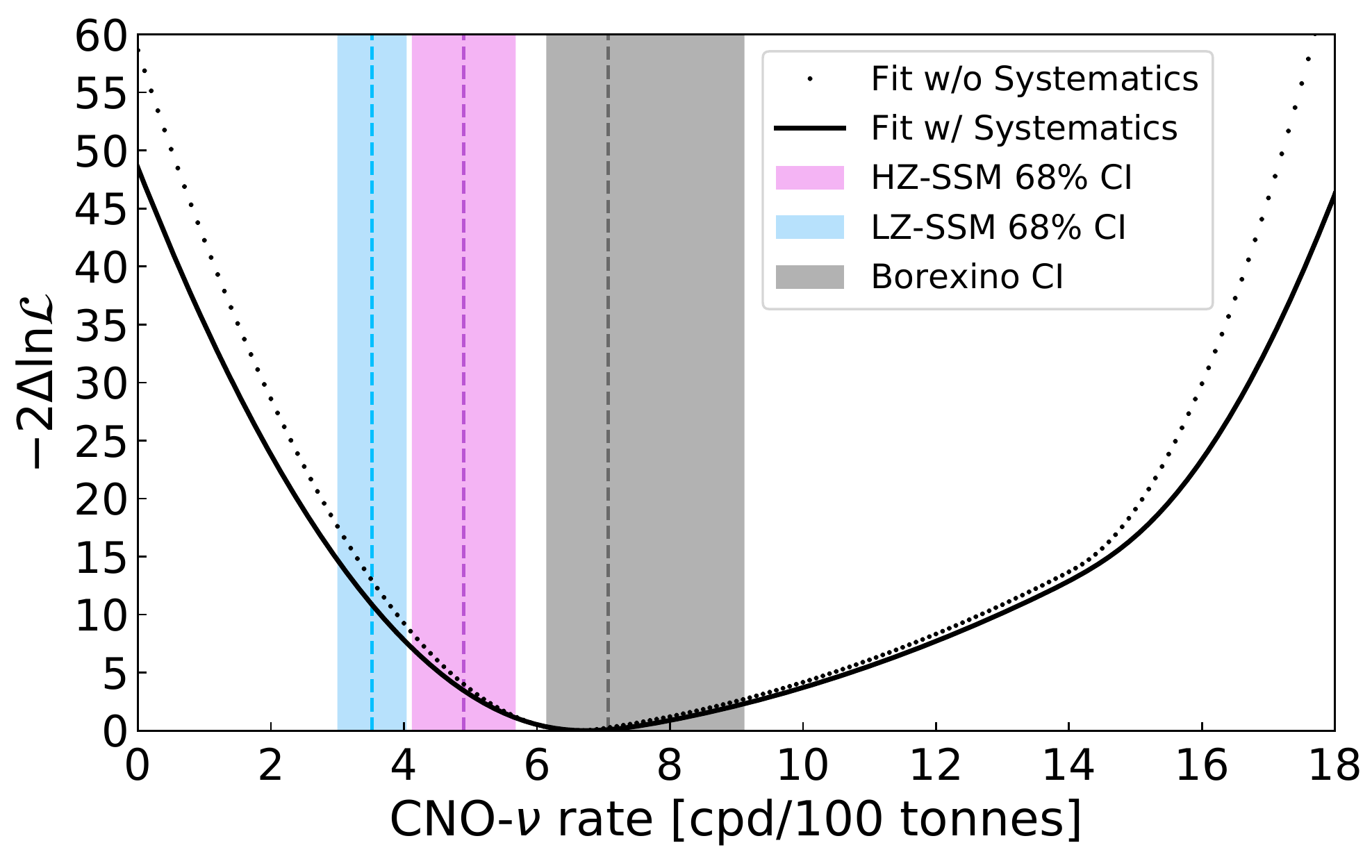}  
    \label{fig:profile}}
    \caption{(a) Spectral fit (magenta) of the Borexino Phase-III data (black points) from January 2017 to October 2021 with a suppressed contribution of the cosmogenic $^{11}$C background (grey dashed). CNO-neutrinos are shown as red solid line. The rates of $pep$ neutrinos (green) and $^{210}$Bi (blue) were constrained in the fit based on independent data. The energy estimator $N_{\mathrm{h}}$, in which the fit is performed, represents the number of detected photoelectrons, normalized to 2,000 live channels. (b) CNO-neutrino rate negative log-likelihood ($-2\Delta \ln \mathcal{L}$) profile obtained from the multivariate spectral fit (dashed black line) and after folding in the systematic uncertainties (black solid line). The blue, violet, and grey vertical bands show 68\% confidence intervals (C.I.) for the low metallicity SSM B16-AGSS09met ((3.52 $\pm$ 0.52)\,cpd/100\,tonnes) and the high metallicity SSM B16-GS98 ((4.92 $\pm$ 0.78)\,cpd/100\,tonnes) predictions~\cite{Vinyoles:2016djt,sensitivity}, and the new Borexino result including systematic uncertainty, respectively.}
\end{figure*}

{\it Borexino and Phase-III dataset} --- Borexino is a large volume liquid scintillator experiment, located at Laboratori Nazionali del Gran Sasso in Italy and has operated from May 2007 until October 2021. The core of the detector~\cite{BorexinoDetector} consists of  $\sim$280\,tonnes of liquid scintillator contained in a 4.25\,m radius, 125\,$\mu$m thick nylon vessel. The concentric detector geometry is designed to shield the innermost scintillator from radioactivity originated from external materials. The scintillation light is detected by nominally 2212 photomultiplier tubes (PMTs) mounted on a 7\,m radius stainless steel sphere (SSS). Since the solar neutrino signal is rare and indistinguishable from natural radioactivity, radiopurity and background control are the pivotal key to success. The underground location reduces the cosmic muon flux by a factor of $\approx$\,10$^6$, while a water Cherenkov veto surrounding the SSS tags residual muons. During the initial filling, the scintillator was purified~\cite{ALIMONTI200958} to unprecedented levels of radiopurity\,\cite{bx-phaseI}, further improved\,\cite{bx-nusol} by operations performed in 2010-2011. 

As discussed in~\cite{CNO-nature,sensitivity}, constraining the decay rate of $^{210}$Bi, a daughter of $^{210}$Pb contaminating the scintillator, is a key requirement for the CNO analysis and is achieved by measuring the $\alpha$ decay rate of the $^{210}$Bi daughter, $^{210}$Po\,\cite{Villante_Po}. This procedure is severely limited by out-of-equilibrium $^{210}$Po in the analysis volume, originating from the vessel surface and carried over by temperature-driven seasonal convective currents. Between 2015 - 2019, the Borexino detector was thermally stabilized to suppress this effect. This made possible the first evidence of CNO $\nu$'s~\cite{CNO-nature} using data collected from July 2016 until February 2020. This paper is based on data taken when the radiopurity and thermal stability of the detector was maximal, $i.e.$, between January 2017 and October 2021 (final Phase-III). The last part of the dataset features an unprecedented thermal stability and an enlarged volume of strongly reduced $^{210}$Po contamination (see Fig.~\ref{fig:210Po_map}), and therefore provides an improved $^{210}$Bi constraint. Furthermore, we now exclude the second half of 2016 used in~\cite{CNO-nature}, as it was still affected by an evident amount of out-of-equilibrium $^{210}$Po. The overall exposure of the analysis presented in this paper is 1431.6\,days\,$\times$\,71.3\,tonnes, 33.5\% more than in~\cite{CNO-nature}.

{\it Analysis strategy and results} --- In Borexino, solar neutrinos are detected via their elastic scattering off electrons. Thus, the detected signal is induced by the electrons characterized by a continuous energy distribution even for mono-energetic neutrinos as $^7$Be or $pep$. For CNO $\nu$'s, produced in an energy interval extending up to 1740\,keV, the electron spectrum is rather featureless with an end-point at 1517\,keV and with a low expected interaction rate of few counts per day (cpd) in 100\,tonnes of scintillator. In order to disentangle the CNO-$\nu$ signal from other solar $\nu$'s and backgrounds, we follow the same procedure applied in~\cite{CNO-nature}. The multivariate fit is performed on two energy spectra from 320\,keV to 2640\,keV and the radial distribution of selected events. The two energy spectra are obtained by dividing the selected events into two complementary datasets, with and without cosmogenic $^{11}$C, using the {\it Three-Fold Coincidence} procedure~\cite{bib:TFC}. All events must be reconstructed in a wall-less, centrally located 71.3\,tonnes fiducial volume. The shapes of all signal and background components are obtained with a full Geant4 based Monte Carlo simulation~\cite{bib:MCPaper}, with an improved treatment of the time evolution of PMT's effective quantum efficiencies based on the low-energy $^{14}$C data. We note that Borexino is not sensitive to the small dependence of the shape of solar neutrino components on neither the oscillation parameters nor the relative ratio of the individual CNO components. Thus, in the Monte Carlo production, we assume the standard 3-flavour neutrino oscillations and the $^{13}$N, $^{15}$O, and $^{17}$F relative contributions to the CNO flux according to SSM B16~\cite{Vinyoles:2016djt}.

 The main part of the sensitivity to CNO~\cite{sensitivity} comes from the  $^{11}$C-depleted spectrum shown in Fig.~\ref{fig:TFCsubtracted}, in which the CNO end-point is ``unveiled" by the removal of about 90\% of $^{11}$C, while preserving more than 60\% of the exposure. Further complications arise from the degeneracy of the CNO energy spectrum with those of $pep$ solar $\nu$'s and $^{210}$Bi. The $pep$ rate is constrained in the fit to the value (2.74\,$\pm$\,0.04)\,cpd/100\,tonnes as in~\cite{CNO-nature}. A constraint on $^{210}$Bi is evaluated from the minimum rate of its daughter $^{210}$Po. Since we cannot exclude small levels of out-of-equilibrium $^{210}$Po from residual convection, we consider this minimum as an upper limit on $^{210}$Bi and implement it as a half-Gaussian penalty term in the likelihood. The $\alpha$ decays of $^{210}$Po are identified on an event-by-event basis using the pulse shape discrimination neural network method~\cite{Hocker:2007ht, CNO-nature}. A {\it Low Polonium Field} (LPoF) volume is identified as the region of the detector with the lowest $^{210}$Po contamination, quantified via a fit with a 2D paraboloid equation (with and without a cubic spline function along the $z$-axis to account for more complexity in this direction) as in~\cite{CNO-nature}. Since the $z$ position of the LPoF is slightly changing in time due to residual convective motions, especially before 2020, we first performed the fits on the monthly LPoF data in an enlarged volume of 70\,tonnes, in order to obtain its positions shown in Fig.~\ref{fig:210Po_map}. These are then used to blindly align monthly datasets using the previous month's position. It should be noted that the LPoF has been extremely stable from August 2020 until the end of data-taking, and has significantly increased in size. The final LPoF fit is then performed on the aligned dataset in 20-25\,tonnes, depending on the method, on approximately 6,000-9,000 $^{210}$Po events. The final $^{210}$Bi upper limit including all systematic uncertainties is (10.8\,$\pm$\,1.0)\,cpd/100\,tonnes. This value is lower, yet compatible with the previous limit of (11.5\,$\pm$\,1.3)\,cpd/100\,tonnes~\cite{CNO-nature}, thanks to the removal of the 2016 data with high $^{210}$Po rate, and more precise due to the inclusion of the new stable period after February 2020. The major systematic contribution of 0.68\,cpd/100\,tonnes is associated with the $^{210}$Bi spatial uniformity in the fiducial volume, a necessary pre-requisite in order to apply the $^{210}$Bi constraint in a volume $\sim$3 times larger than the LPoF. This error has been estimated independently by studying $\beta$-like events in the energy region with maximum relative contribution of $^{210}$Bi, in the the entire fiducial volume, and split into radial and angular components, as in~\cite{CNO-nature}. 
The final fit with the $pep$ and $^{210}$Bi rates constrained is shown in Fig.~\ref{fig:TFCsubtracted} on the $^{11}$C-subtracted energy spectrum. The rates of additional backgrounds, i.e., the external $\gamma$'s from $^{40}$K, $^{208}$Tl, and $^{214}$Bi, $^{85}$Kr and $^{210}$Po in the scintillator, cosmogenic $^{11}$C, as well as $^{7}$Be solar $\nu$'s are kept as free fit parameters. The model fits to the data with a $p$-value of 0.2 and yields the CNO-$\nu$ interaction rate with zero threshold of
$6.6^{+2.0}_{-0.7}$\,cpd/100\,tonnes. The corresponding negative log-likelihood profile for the CNO-$\nu$ rate, shown in dashed-line in Fig.~\ref{fig:profile}, is asymmetric since the upper limit $^{210}$Bi constraint impacts only the left part of the CNO profile. The right part of the CNO profile is unconstrained by the penalty and exploits the small difference between the CNO and $^{210}$Bi spectral shapes. The solid line in Fig.~\ref{fig:profile} shows the CNO profile including the total systematic uncertainty of $^{+0.5}_{-0.4}$\,cpd/100\,tonnes, evaluated with the same toy-Monte-Carlo-based method as in~\cite{CNO-nature}. The extent of individual parameters left to vary in this procedure has been updated for the current analyzed period, using improved Monte Carlo simulations and 2.2\,MeV $\gamma$'s from the cosmogenic neutron capture on scintillator hydrogen (instead of $\alpha$-decays from non-homogeneously distributed  $^{210}$Po used in~\cite{CNO-nature}) as a standard candle for the detector stability and uniformity. The final result on the CNO-$\nu$ interaction rate with zero-threshold is $6.7^{+2.0}_{-0.8}$\,cpd/100\,tonnes, obtained from the 68\% quantile of the likelihood profile including the systematic uncertainty. This result excludes the no-CNO-signal hypothesis at about 7$\sigma$ C.L. Taking into account the density of electrons in the scintillator of (3.307\,$\pm$\,0.015)\,$\times 10^{31}$\,$e^-$/100\,tonnes and assuming Mikheyev-Smirnov-Wolfenstein flavour conversion in matter~\cite{PhysRevD.17.2369, Mikheev:1986gs,deHolanda2004} and the neutrino oscillation parameters from~\cite{Esteban:2020cvm}, the measured rate including systematic uncertainty is converted into a flux of $6.6^{+2.0}_{-0.9} \times 10^8$\,cm$^{-2}$\,s$^{-1}$ CNO solar $\nu$'s on Earth. 

\begin{figure}
\centering
    \includegraphics[width=0.46\textwidth]{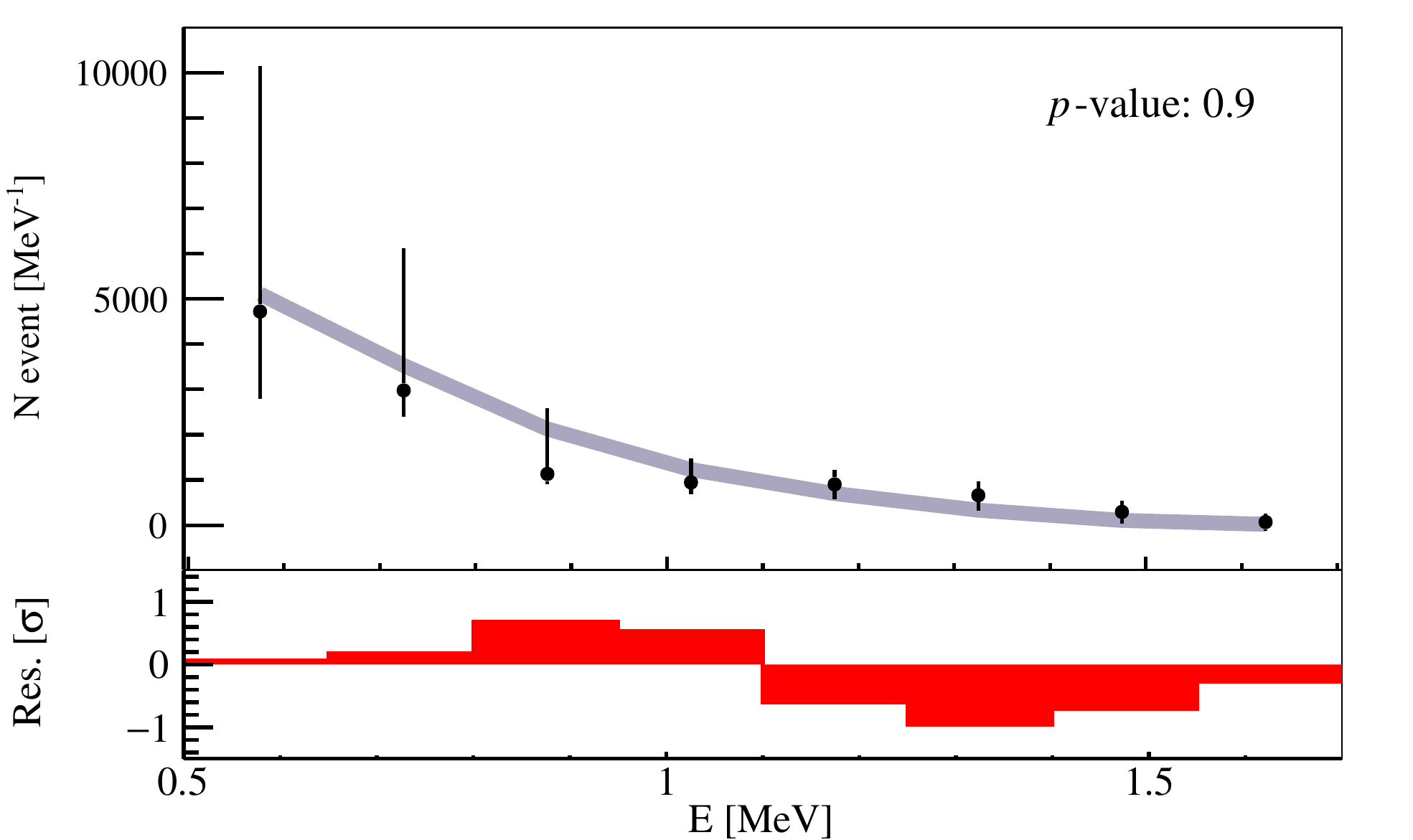}
    \caption{Top: spectral shape of the events after subtraction of all known backgrounds (black dots). The gray line is the fitted Monte-Carlo-based CNO shape assuming standard neutrino interaction and oscillation. Bottom: residual (Res.) of the fit, defined as (model-data)$/\sigma_{\mathrm{data}}$, shows the data compatibility with the expected shape of recoiled electrons from CNO $\nu$'s.}
    \label{fig:CNOspectrum}
\end{figure}

 We have tested whether the events in excess to all known backgrounds, determined excluding the CNO $\nu$ energy range, are compatible with the expected CNO energy spectrum.
The rate of the external background and cosmogenic $^{11}$C is obtained by the multivariate fit of events' energy and radial distributions above the CNO end-point.
The $^{85}$Kr background is evaluated using the fast coincidence tagging method~\cite{bx-phaseI}, not used in the main analysis.
The rate of $^7$Be solar $\nu$'s is taken from the Borexino Phase-II results~\cite{ppchain-nature}. The $^{210}$Po rate is obtained by fitting $\alpha$-like events selected by $\alpha/\beta$ discrimination methods. The rate of $pep$ $\nu$'s is set to the value of the constraint used in the main analysis. For the $^{210}$Bi we subtract the asymmetric value of $10.8^{+1.0}_{-10.8}$\,cpd/100\,tonnes motivated by our upper limit constraint. The energy distribution of events 
after subtracting all the background contributions, shown in Fig.\,\ref{fig:CNOspectrum}, is
found compatible with the CNO expected shape ($p$-value of 0.9).

\begin{figure}[t]
    \centering
    \includegraphics[width=0.47\textwidth]{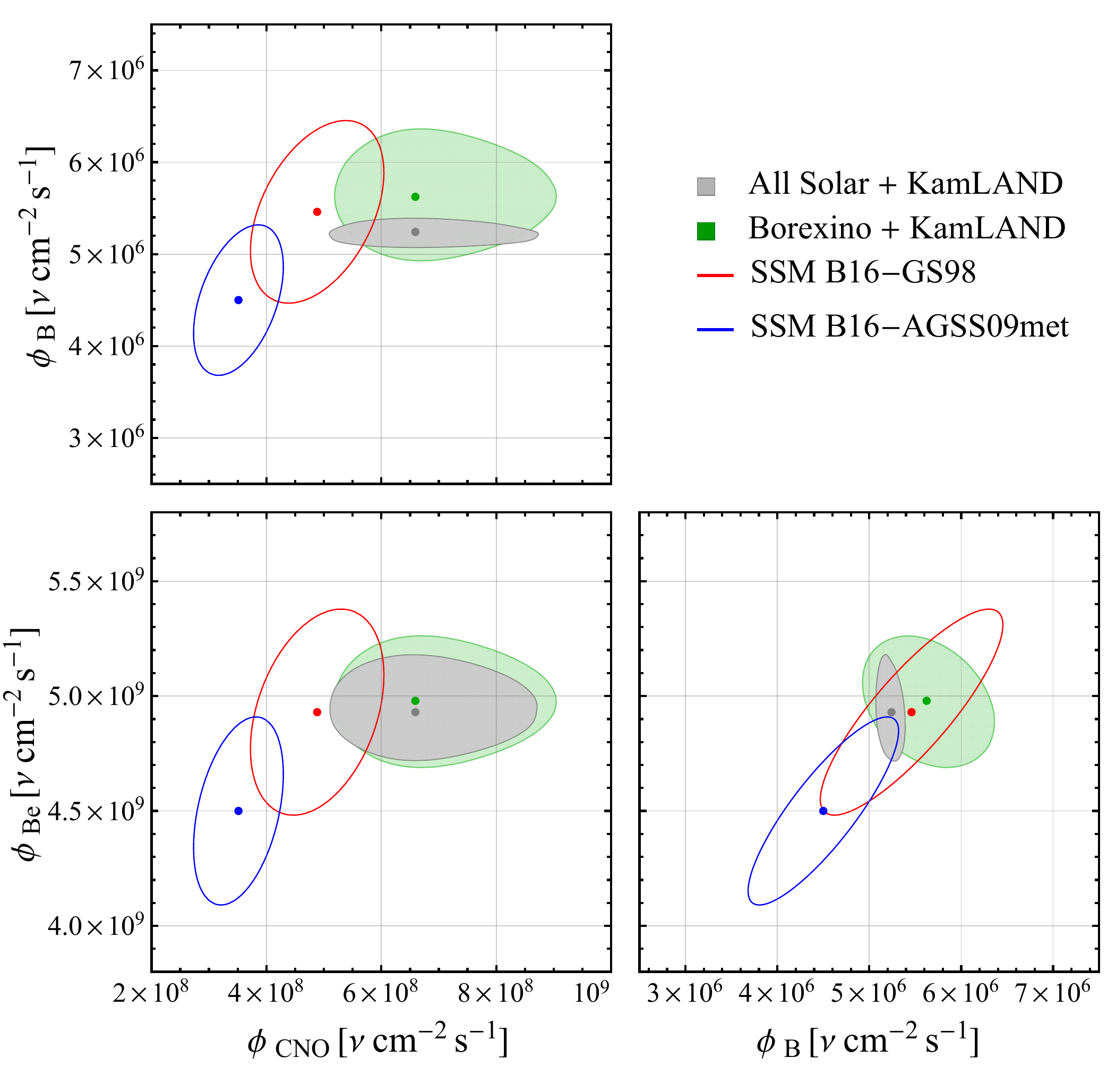}
    \vspace{0.3em}
    \caption{Results of the global analysis of solar neutrino and KamLAND reactor data (grey regions) and of Borexino only + KamLAND (green regions) in the $\Phi_{\ce{B}}$-$\Phi_{\ce{Be}}$, 
$\Phi_{\ce{B}}$-$\Phi_{\ce{CNO}}$, and $\Phi_{\ce{Be}}$-$\Phi_{\ce{CNO}}$ 
planes. The predictions of high-metallicity SSM B16-GS98 model (red) and low-metallicity SSM B16-AGSS09met (blue) are also shown.
 The best fit values of $\Delta m^2_{21}$\,=\,$7.50^{+0.17}_{-0.18}\times 10^{-5}$\,eV$^2$ and $\tan \theta_{12}$\,=\,$0.43^{+0.04}_{-0.02}$.}
    \label{fig:fit_a}
\end{figure}

{\it Implications for solar physics} ---
We perform a global analysis of all solar $\nu$ data  to test their compatibility with the
SSM B16 predictions on solar neutrino fluxes~\cite{Vinyoles:2016djt}. We follow the procedure discussed in~\cite{bx-phaseI, ppchain-nature} and include, together with the new CNO rate measurement, also the data
from radiochemical experiments~\cite{Cleveland:1998nv, SAGE:2009eeu, Kaether:2010ag}, $^8$B-$\nu$ data from SNO~\cite{SNO:I, SNO:II} and Super-Kamiokande~\cite{SK2006, SK2016}, and Borexino Phase\,II~\cite{ppchain-nature} results on $^7$Be and $^8$B $\nu$'s, as well as the KamLAND reactor $\bar{\nu}_e$ data~\cite{KamLAND:2010fvi} to better constrain $\Delta m^{2}_{21}$. The fluxes $\Phi$ of \ce{^{8}B}, \ce{^{7}Be}, and \ce{CNO} $\nu$'s, as well as
$\Delta m^{2}_{12}$ and $\theta_{12}$ are left free in the fit, while $\theta_{13}$,
having a negligible impact in the analysis, has been fixed according to\,\cite{Esteban:2020cvm}.
The results are shown in Fig.\,\ref{fig:fit_a}, where the grey areas are the 1$\sigma$ allowed regions in the $\Phi_{\ce{B}}$-$\Phi_{\ce{Be}}$,
$\Phi_{\ce{B}}$-$\Phi_{\ce{CNO}}$, and $\Phi_{\ce{Be}}$-$\Phi_{\ce{CNO}}$ 
planes. We also display the output of the fit when only results from Borexino and KamLAND are included (green areas). The predictions of the SSM B16 are represented by the elliptical contours, when the high-metallicity GS98 (red) and low-metallicity AGSS09met (blue) inputs are used. It is clear that both results exhibit a small tension with SSM B16-AGSS09met prediction, that is driven by the CNO $\nu$'s. We quantify the tension using the test-statistics introduced in~\cite{MaltoniPG}. We find that the $p$-value of the comparison between the low-metallicity SSM B16-AGSS09met predictions and the global analysis results worsens from 0.327 to 0.028, when including the CNO measurement. The same happens in the comparison with Borexino-only data, where the $p$-value lowers from 0.196 to 0.018 when including CNO. On the other hand, the high-metallicity SSM B16-GS98 is fully compatible with both the global analysis and the Borexino-only results in all cases ($p$-value\,=\,0.462 and 0.554 including CNO, respectively). Following the procedure described in~\cite{ppchain-nature}, we also performed a frequentist hypothesis test based on a likelihood-ratio test statistics including only Borexino results on $^7$Be, $^8$B, and CNO $\nu$'s. Assuming SSM B16-GS98, our data disfavors SSM B16-AGSS09met at 3.1$\sigma$ ($p$-$value$\,=\, 9.1$\cdot$ 10$^{-4}$).

The interpretation of the observed tension between data and SSM B16-AGSS09met predictions is non univocal due to the degeneracy between metallicity, opacity, and other inputs of the SSM.
More information on metallicity can be gathered by exploiting the direct dependence of the CNO cycle from the C and N abundances in the core of the Sun, in combination with the precise measurement of the $^8$B-$\nu$ flux, as suggested in~\cite{Haxton:2008, Serenelli:2013} and discussed specifically for Borexino in~\cite{sensitivity}.
The general idea of this method is the following: solar neutrino fluxes (both those produced in the $pp$ chain and in the CNO cycle) depend on the so-called {\it environmental} parameters (abundances of heavy elements, solar age, luminosity, opacity, diffusion) only indirectly, through the core temperature \Tc{}, which is an implicit function of them. Therefore, the uncertainties affecting these parameters  collapse into the overall uncertainty of the temperature profile. The dependence of the neutrino flux $\Phi_i$ on \Tc{} can be approximated by a power-law, with power index $\tau_i$ specific to the flux under consideration. The flux of \ce{^{8}B} $\nu$'s is the most sensitive to variations of \Tc{}, 
featuring a power index $\tau_{\ce{B}} \approx 24$ \cite{VillanteSerenelli_XTemp}, 
\begin{equation}
\FluxB{}/\FluxB\ssm{} \propto (\Tc{}/\Tc\ssm{})^{\tau_{\ce{B}}},
\label{eq:B}
\end{equation}
with ``SSM" indicating the SSM predicted value.
The same relationship holds for reactions belonging to the CNO cycle, like for example $^{15}$O, but with a different exponent $\tau_{\ce{O}} \approx 20$\,\cite{VillanteSerenelli_XTemp}.
In addition, CNO reactions' rate feature a direct dependence on the 
abundance of \ce{C} and \ce{N} (relative to hydrogen) in the solar core
$n_{\ce{CN}} = (n_{\ce{C}} + n_{\ce{N}}$):
\begin{equation}
    \Phi_{\ce{O}}/\Phi^\text{SSM}_{\ce{O}} \propto 
    \frac{n_{\ce{CN}}}{n_{\ce{CN}}^\text{SSM}} \times
    (\Tc{}/\Tc\ssm{})^{\tau_{\ce{O}}}.
    \label{eq:O}
\end{equation}
It is then possible to construct a weighted ratio between the \ce{^{15}O} and 
\ce{^{8}B} fluxes of the form:
\begin{equation}
    (\Phi_{\ce{O}}/\Phi_{\ce{O}}^\text{SSM})/(\Phi_{\ce{B}}/\Phi^\text{SSM}_{\ce{B}})^k,
    \label{eq:ratio}
\end{equation}
with $k$ chosen to minimize the impact of \Tc{} (and therefore of its uncertainties) on the ratio,
thus isolating the effect of $n_{\ce{CN}}/n_{\ce{CN}}^\text{SSM}$. 
Substituting Eq.\,\eqref{eq:B} and Eq.\,\eqref{eq:O} in Eq.\,\eqref{eq:ratio}, we obtain
\begin{equation}
 \frac{(\FluxO{}/\FluxO\ssm{})}{(\FluxB{}/\FluxB\ssm{})^k} \propto
 \frac{n_{\ce{CN}}}{n_{\ce{CN}}\ssm{}}\,\left(\frac{\Tc{}}{\Tc\ssm{}} \right)^{\tau_{\ce{O}} - k\tau_{\ce{B}}}
 \label{eq:wr}
\end{equation}
and the appropriate value of $k$ would therefore be $\tau_{\ce{O}}/\tau_{\ce{B}}$= 0.83. The above equation is, however, oversimplified since both \ce{^{8}B} and \ce{^{15}O} $\nu$'s are produced in a relatively small region of the solar core, where the temperature and the chemical composition vary. 
In addition, both the temperature and the composition profile are non-trivial functions of the SSM input parameters. 
    
To overcome this hurdle, we must explicit the dependence of the $^8$B and $^{15}$O fluxes on each SSM input parameter, making use of the corresponding partial logarithmic derivatives, following~\cite{Bahcall:1989ks, Haxton:2008, Serenelli:2013}. Taking the SSM B16-GS98 model as a reference, we obtain that $k$ = 0.769 minimizes the impact of the environmental parameters on the fluxes ratio in Eq.~\ref{eq:ratio} (more details in Supplemental Material).  With this optimized value of $k$, we find:
    \begin{multline}
    \frac{(\Phi_{\ce{O}}/\Phi_{\ce{O}}^\text{SSM})}{(\Phi_{\ce{B}}/\Phi^\text{SSM}_{\ce{B}})^{0.769}} =
    \frac{N_{\ce{CN}}}{N_{\ce{CN}}^\text{SSM}}
    \times \\
    \times \left[1 \pm (0.097\text{(nucl)} \oplus 0.005\text{(env)} \oplus 0.027\text{(diff)}) \right].
\label{eq:rk}
\end{multline}
The terms in square brackets quantify the contributions of the nuclear, environmental, and diffusion  uncertainties to the error budget to be summed in quadrature. Note that, the symbol $N_{\mathrm{CN}}$ represents the C\,+\,N abundance in the photosphere and not in the solar core. Indeed, the partial derivatives used in this procedure~\cite{B16SSMonline} are evaluated with respect to the composition of the photosphere, where spectroscopic data provide observational constraints. 

\begin{figure}
    \centering
    \includegraphics[width=0.95\columnwidth]{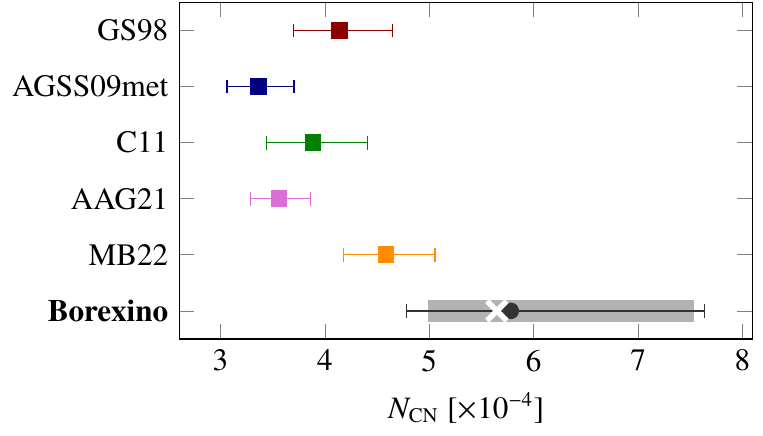}  
    \caption{
        Comparison of abundance of (C\,+\,N)$/$H in the solar photosphere, $N_{\ce{CN}}$, from spectroscopy (squares) and from the Borexino measurement (circle). The gray area highlights the uncertainty due to the precision of the 
        CNO rate measurement. The white cross marks the result of the very same
        study repeated changing the reference SSM from the B16-GS98
        to the B16-AGSS09met. 
    }
    \label{fig:forest} 
\end{figure}

Inserting in Eq.\,\ref{eq:rk} the flux of \ce{^{8}B} $\nu$'s obtained from the global analysis ($\Phi_{\ce{B}}/\Phi_{\ce{B}}^\text{SSM}$\,=\,0.96\,$\pm$\,0.03) and $\Phi_{\ce{O}}/\Phi_{\ce{O}}^\text{SSM}=1.35^{+0.41}_{-0.18}$ extracted from our CNO measurement, assuming the SSM ratio between \ce{^{13}N} and
\ce{^{15}O} fluxes, we obtain:
\begin{multline}
    \frac{N_{\ce{CN}}}{N_{\ce{CN}}^\text{SSM}} = 
    1.35 \times \left(0.96\right)^{-0.769}\times \\
    \times\left[ 
      1 \pm \left(_{-0.136}^{+0.303}(\ce{CNO}) \oplus 0.097(\text{nucl}) \oplus 0.023(\ce{^{8}B}) \right.\right. \\
      \left.\left. \oplus   \,0.005(\text{env}) \oplus 0.027(\text{diff}) \oplus 0.022(\ce{^{13}N}/\ce{^{15}O})\right) \right].
    \label{eq:final}
\end{multline}
By construction, the contribution to the error budget from environmental variables is negligible, while the precision 
of the $R_{\ce{CNO}}$ measurement is dominant. The leading  residual uncertainty of $9.7\%$ comes from the astrophysical $S$-factors, driven by $S_{114}$ ($7.8\%$) and $S_{17}$ ($3.7\%$). The error on the extrapolation of the C\,+\,N abundance from the core to the photosphere due to diffusion is 2.7\%.
Finally, the C\,+\,N abundance with respect to the H in the photospehere is 
$N_{\ce{CN}} = (5.78^{+1.86}_{-1.00})\times 10^{-4}$.
This represents the first determination of the abundance of C\,+\,N in the Sun using neutrinos.
Our result is compared to the measurements based on spectroscopy of the photosphere in
Fig.\,\ref{fig:forest}. It is in good agreement with the recent MB22~\cite{MB22} and the outdated GS98~\cite{GS98}  compilations,
while it shows a moderate $\sim$2$\sigma$ tension with the values of AGSS09met~\cite{AGSS09,SMaccretion} and 
its recent update AAG21~\cite{AAG21}. 
The stability of our result with respect to the input metallicity is demonstrated by repeating the analysis changing our reference to SSM B16-AGSS09met and obtaining fully compatible value (white cross in Fig.\,\ref{fig:forest}).

{\it Outlook} --- In this letter, we have presented the latest Borexino measurement of the CNO solar $\nu$'s with an improved uncertainty of $_{-12\%}^{+30\%}$ on its rate.
This result reinforces the one previously published by Borexino in 2020~\cite{CNO-nature}, now further increasing the detection significance to about 7$\sigma$ C.L. against the null hypothesis. 
We included this new result in the global analysis of all solar $\nu$ and KamLAND reactor data. We found the resulting solar $\nu$ fluxes to be in agreement with the ``high metallicity" SSM B16-GS98~\cite{Vinyoles:2016djt}, while a moderate tension is observed when ``low metallicity" AGSS09met is used for the SSM prediction. 
A frequentist hypothesis test using only Borexino CNO, $^7$Be, and $^8$B $\nu$'s fluxes, disfavors the SSM B16-AGSS09met at 3.1$\sigma$\,C.L. as an alternative to SSM B16-GS98.
In addition, we have used the CNO $\nu$ measurement together with the $^8$B $\nu$ flux from the global analysis to determine the C\,+\,N abundance in the Sun, breaking the ambiguity due to the opacity/metallicity degeneracy. The C\,+\,N abundance determined with this method, was compared  with the independent spectroscopic measurements of the solar photosphere. Even though affected by a large error of $_{-17\%}^{+32\%}$ (dominated by the error on the measured CNO rate) our measurement agrees very well with the so-called high-metallicity compilations (MB22\,\cite{MB22}, GS98\,\cite{GS98}), while featuring a moderate $\sim$2$\sigma$ tension with the low-metallicity ones (AGSS09met~\cite{AGSS09,SMaccretion}, C11\,\cite{C11}, AAG21\,\cite{AAG21}). A more precise measurement of the CNO flux, performed by future experiments could provide an important element to definitively assess the long standing metallicity controversy and to constrain the range of possible non-standard solar models~\cite{SMaccretion,SMcvovershoot}.

{\it Acknowledgments:}
We acknowledge the generous hospitality and support of the Laboratori Nazionali del Gran Sasso (Italy). The Borexino program is made possible by funding from Istituto Nazionale di Fisica Nucleare (INFN) (Italy), National Science Foundation (NSF) (USA), Deutsche Forschungsgemeinschaft (DFG), Cluster of Excellence PRISMA+ (Project ID 39083149), and recruitment initiative of Helmholtz-Gemeinschaft (HGF) (Germany), Russian Foundation for Basic Research (RFBR) (Grants No. 19-02-00097A) and Russian Science Foundation (RSF) (Grant No. 21-12-00063) (Russia), and Narodowe Centrum Nauki (NCN) (Grant No. UMO 2017/26/M/ST2/00915) (Poland). We gratefully acknowledge the computing services of Bologna INFN-CNAF data centre and U-Lite Computing Center and Network Service at LNGS (Italy). FLV acknowledges support by ``Neutrino and Astroparticle Theory Network'' under the programme PRIN 2017 funded by the Italian Ministry of Education, University and Research (MIUR) and INFN Iniziativa Specifica TAsP.

\bibliographystyle{apsrev4-1}
%
 

\clearpage
\appendix
\noindent
\begin{center}
\textbf{Supplemental Material}
\end{center}

In this appendix we provide further details about the method used to obtain  the abundance of carbon and nitrogen from the new CNO neutrino flux measurement presented in this letter.

The concept of this procedure, first proposed in \cite{Haxton:2008, Serenelli:2013}, is to use the $^8$B neutrino flux measurement as a ``thermometer'' to constrain the temperature of the solar core. In this way, the temperature dependence of the CNO neutrino flux can be removed, making it possible 
to exploit the direct connection between the power produced by the CNO cycle and the abundance of 
carbon and nitrogen in the core to determine the latter from a measurement of the 
CNO neutrino flux. 
In practice, we can achieve this result by constructing a weighted ratio between 
one of the neutrino fluxes generated in the CNO cycle (such as the one of \ce{^{15}O},
$\FluxO$) and the flux of \ce{^{8}B} neutrinos $\FluxB$ 
(the most sensitive probe of temperature deviations in the solar core), with a proper weighting factor $k$ which is chosen to minimize variations due to temperature.

Approximating the relationship between solar 
neutrino fluxes and variations in the solar core temperature \Tc{} with a power-law 
\cite{VillanteSerenelli_XTemp}, we write
\begin{equation}
    \frac{(\FluxO/\FluxO\ssm{})}{(\FluxB{}/\FluxB\ssm{})^k} \propto
    \frac{n_{\ce{CN}}}{n_{\ce{CN}}\ssm{}}\,\left(\frac{\Tc{}}{\Tc\ssm{}} \right)^{\tau_{\ce{O}} - k\tau_{\ce{B}}},
    \label{eq:sm_wr}
\end{equation}
where $\tau_{\ce{B}(\ce{O})} \approx 24(20)$, $n_{\ce{CN}}$ denotes the abundance of 
carbon and nitrogen relative to hydrogen in the solar core, and the label ``SSM'' 
indicates the SSM predicted value.

As discussed in the main text, Eq.\,\eqref{eq:sm_wr} cannot be used directly to access
$n_{\ce{CN}}$. First, neutrinos from \ce{^{8}B} and \ce{^{15}O} are produced in 
an extended region of the solar core where both the temperature and chemical 
composition profiles vary; second, the temperature and the C+N abundance profiles 
are not direct inputs of the SSM. The core temperature profile, which in this approximated picture 
reduces to a single value \Tc{}, is indeed a function of a subset of the SSM parameters, 
the so-called \textit{environmental parameters}. These parameters include the astrophysical properties of the Sun 
(i.e., solar age, luminosity $L_\odot$), the description of the solar opacity\footnote{
    The radiative opacity is represented by two parameters, namely $\kappa_a$ 
    and $\kappa_b$, which describes the variation of the solar opacity profile 
    as discussed in\,\cite{Vinyoles:2016djt}.
} ($\kappa$), 
the diffusion parameter and the abundances of heavy elements relative to hydrogen
(\ce{C}, \ce{N}, \ce{O}, \ce{F}, \ce{Ne}, \ce{Mg}, \ce{Si}, \ce{S}, \ce{Ar} and \ce{Fe}), 
which are calibrated according to spectral analyses of the photosphere (often combined
with meteoritic abundances). 

To properly account for the contribution of the SSM parameters in the weighted ratio
of neutrino fluxes in Eq.\,\eqref{eq:sm_wr} we follow the conventional expansion
of the SSM flux predictions \cite{Bahcall:1989ks, Haxton:2008, Serenelli:2013}, 
which makes explicit the dependence of a given neutrino flux $\Phi_i$ from the input $j$
in the form of a power-law
\begin{multline}
    \frac{\Phi_i}{\Phi_i^\text{SSM}} = \prod_j^\text{C,N} x_j^{\alpha(i, j)}
    \times \prod_j^\text{env} x_j^{\alpha(i, j)} \times \prod_j^\text{nucl} x_j^{\alpha(i, j)} 
    \times x_\text{diff}^{\alpha(i, \text{diff})} 
    \label{eq:sm_alpha}
\end{multline}
where $x_j$ denotes the SSM parameters normalized for their nominal values and
the coefficients $\alpha(i, j)$ are the logarithmic derivatives
\cite{Vinyoles:2016djt} 
of the neutrino flux $\Phi_i$ with respect to the SSM parameter $j$ 
\begin{equation}
    \alpha(i, j) = \frac{\partial \ln\left(\Phi_i/\Phi_i^\text{SSM}\right)}{\partial \ln x_j},
    \label{eq:sm_alpha_def}
\end{equation}
which are calculated numerically and given in~\cite{B16SSMonline}.
We notice that the logarithmic derivatives for the composition parameters are 
evaluated by studying the effect of modification of the surface composition on the flux $\Phi_i$
within the range allowed by the observational constraints. 

In Eq.\,\eqref{eq:sm_alpha} the SSM inputs are conveniently grouped into four 
categories: along with the nuclear reaction cross section, we have separated the 
abundances of carbon and nitrogen that are the target of our study. 
The diffusion parameter is also stripped from the environmental parameters 
account because it features a twofold effect: on one hand, a change in the diffusion 
will affect the temperature stratification in the Sun; on the other hand, it will
also affect the chemical composition profile.

Using Eq.\,\eqref{eq:sm_alpha} it is then possible to express the 
weighted ratio on the left-hand side of Eq.\,\eqref{eq:sm_wr} 
as a function of the SSM input parameters
\begin{multline}
    \frac{(\FluxO/\FluxO\ssm{})}{(\FluxB{}/\FluxB\ssm{})^k} = 
    \prod_j^\text{C,N} x_j^{\alpha(\ce{^{15}O}, j) - k\,\alpha(\ce{^{8}B}, j)} 
    \times \prod_j^\text{env} x_j^{\alpha(\ce{^{15}O}, j) - k\,\alpha(\ce{^{8}B}, j)} \times \\
    \times \prod_j^\text{nucl} x_j^{\alpha(\ce{^{15}O}, j) - k\,\alpha(\ce{^{8}B}, j)} 
    \times x_\text{diff}^{\alpha(\ce{^{15}O}, \text{diff}) - k\,\alpha(\ce{^{8}B}, \text{diff})} 
    \label{eq:sm_wr_alpha}
\end{multline}

The optimal value of $k$ is chosen to minimize the contribution of the environmental 
parameters to the total uncertainty budget in the flux ratio in Eq.\,\eqref{eq:sm_wr_alpha}, 
thus making it stable against large variations in the description of the solar temperature 
profile caused by deviations from the assumed chemical composition and/or unconsidered effects in the 
computation of the radiative opacity.
The contribution of uncertainties in the environmental parameters to the variance is
\begin{equation}
    \text{Var}\left[
    \frac{(\FluxO/\FluxO\ssm)}{(\FluxB/\FluxB\ssm)^k} 
    \right]^\text{env} = 
    \sum_{j}^\text{env}\left[ \alpha(\ce{^{15}O}, j) - k\alpha(\ce{^{8}B}, j) \right]^2(\delta x_j)^2,
    \label{eq:var_env} 
\end{equation}
where $\delta x_j$ indicates the fractional uncertainty of the $j$-th environmental parameter.
We assumed for the model inputs the same uncertainties $\delta x_j$ adopted in the
SSM B16~\cite{Vinyoles:2016djt}. For what concerns the chemical abundances, 
which are among the most controversial ingredients of the SSM, 
we have added to the uncertainty an additional contribution to account for the difference between the
GS98 and AGSS09 values (see\,\cite{Serenelli:2013}). 
We choose to use SSM B16-GS98 as reference, although by construction
Eqs.\,\eqref{eq:sm_wr_alpha} stands for any SSM.

Minimizing the variance term in Eq.\,\eqref{eq:var_env}, 
we find the optimal value for $k$ to be 0.769, 
not too far from the value $\tau_{\ce{O}}/\tau_{\ce{B}}\approx0.83$ calculated 
in the simplified picture discussed in the letter.
Substituting $k$\,=\,0.769 in Eq.~(\ref{eq:sm_wr_alpha}) and 
using the tabulated values of the $\alpha(i, j)$
coefficients we obtain
\begin{equation}
    \begin{split}
   \frac{(\FluxO/\FluxO\ssm)}{(\FluxB/\FluxB\ssm)^{0.769}} =  \\
       &\hspace*{-55pt}x_{\ce{C}}^{0.802}                                     
       x_{\ce{N}}^{0.204}                                      
       x_{D}^{0.181}\\
       &\hspace*{-55pt}\times\left[
       x_{S_{11} }^{-0.866}                                  
       x_{S_{33} }^{0.345}                                   
       x_{S_{34} }^{-0.689}                                  
       x_{S_{e7} }^{0.769}                                   
       x_{S_{17} }^{-0.791}                                  
       x_{S_{hep}}^{0.000}                                  
       x_{S_{114}}^{1.046}                                  
       x_{S_{116}}^{0.001}                                          
       \right] \quad \text{(nucl)}\\
       &\hspace*{-55pt}\times \left[ 
       x_\text{Age}   ^{   0.313 }                                  
       x_{L_{\odot}}  ^{   0.602 }                                 
       x_{\kappa_{a}} ^{   0.018 }                                
       x_{\kappa_{b}} ^{  -0.050 }                                       
       \right]\quad\text{(env - solar)}\\
       &\hspace*{-55pt}\times\left[ 
       x_{\ce{O }}^{  0.006}                                   
       x_{\ce{Ne}}^{ -0.003}                                   
       x_{\ce{Mg}}^{ -0.003}                                   
       x_{\ce{Si}}^{  0.001}                                   
       x_{\ce{S }}^{  0.001}                                   
       x_{\ce{Ar}}^{  0.001}                                   
       x_{\ce{Fe}}^{  0.005}           
       \right] \quad \text{(env - met)}
    \end{split}
    \label{eq:sm_breakdown}
\end{equation}
where the first line indicates the contribution of the carbon and nitrogen abundances along with 
the diffusion parameter, the second line highlights the different impact of nuclear cross sections, 
while finally the third and fourth line show the remaining effect of the parameters acting on the 
temperature profile. 

We note that indeed the above relationship features a linear dependence upon 
the abundance of carbon and nitrogen, as it was assumed in the simplified 
description in the main text on the basis of an intuitive argument: 
as discussed in \cite{Serenelli:2013, sensitivity}, when the 
power indices of $x_{\ce{C}}$ and $x_{\ce{N}}$ sum up to one ($0.802 + 0.204 \approx 1$)
one can replace $x_{\ce{C}}^{0.802}\,x_{\ce{N}}^{0.204}$ with the ratio between 
the \ce{C}\,+\,\ce{N} abundance and its nominal value $N_{\ce{CN}}/N_{\ce{CN}}\ssm$.

It should be noticed that in the original works\,\cite{Haxton:2008, Serenelli:2013}, 
this procedure has been proposed to probe the \textit{primordial} C\,+\,N abundance. 
This quantity is however proportional to the surface C\,+\,N abundance apart 
from modification of the efficiency of elemental diffusion, which
are considered in the overall error budget.
As a consequence, we can consider $N_{\ce{CN}}$ as referred to the photosphere
and use this approach to test the surface abundance of C\,+\,N, 
which allow us to make a direct comparison between 
our result and the ones of spectroscopic analysis of the photosphere. 

Using the uncertainties of the SSM inputs discussed above, we can then 
estimate the error budget in Eq.\,\ref{eq:sm_breakdown}, which results in
\begin{multline}
    \frac{(\Phi_{\ce{O}}/\Phi_{\ce{O}}^\text{SSM})}{(\Phi_{\ce{B}}/\Phi^\text{SSM}_{\ce{B}})^{0.769}} =
    \frac{N_{\ce{CN}}}{N_{\ce{CN}}^\text{SSM}}
    \times \\
    \times \left[1 \pm 0.097\text{(nucl)} \pm 0.005\text{(env)} \pm 0.027\text{(diff)} \right].
\label{eq:sm_rk}
\end{multline}

It is then natural to use the experimental determination of the \ce{^{8}B} and 
\ce{^{15}O} neutrino fluxes to invert the above equation and estimate the carbon 
nitrogen abundance. 
As discussed in the main text, we use \FluxB{} as obtained from a global analysis of 
solar neutrino data ($\FluxB{}/\FluxB{}\ssm{}\,=\,0.96\pm0.03$), 
and we extract \FluxO{} from our measurement of the CNO neutrinos interaction rate 
($\FluxO/\FluxO\ssm\,=\,1.35^{+0.41}_{-0.18}$)
assuming the ratio between the \ce{^{13}N} and \ce{^{15}O} neutrinos 
predicted by the SSM and propagating the uncertainty to the final result, 
which yields
\begin{multline}
    \frac{N_{\ce{CN}}}{N_{\ce{CN}}^\text{SSM}} = 
    1.35 \times \left(0.96\right)^{-0.769}\times \\
    \times\left[ 
      1_{-0.136}^{+0.303}(\ce{CNO}) 
      \pm 0.097(\text{nucl}) \pm 0.023(\ce{^{8}B}) \right. \\
      \left. \pm 0.005(\text{env}) \pm 0.027(\text{diff}) \pm 0.022(\ce{^{13}N}/\ce{^{15}O})
    \right].
    \label{eq:sm_final} 
\end{multline}

\begin{figure}
    \centering
    \includegraphics[width=.9\columnwidth]{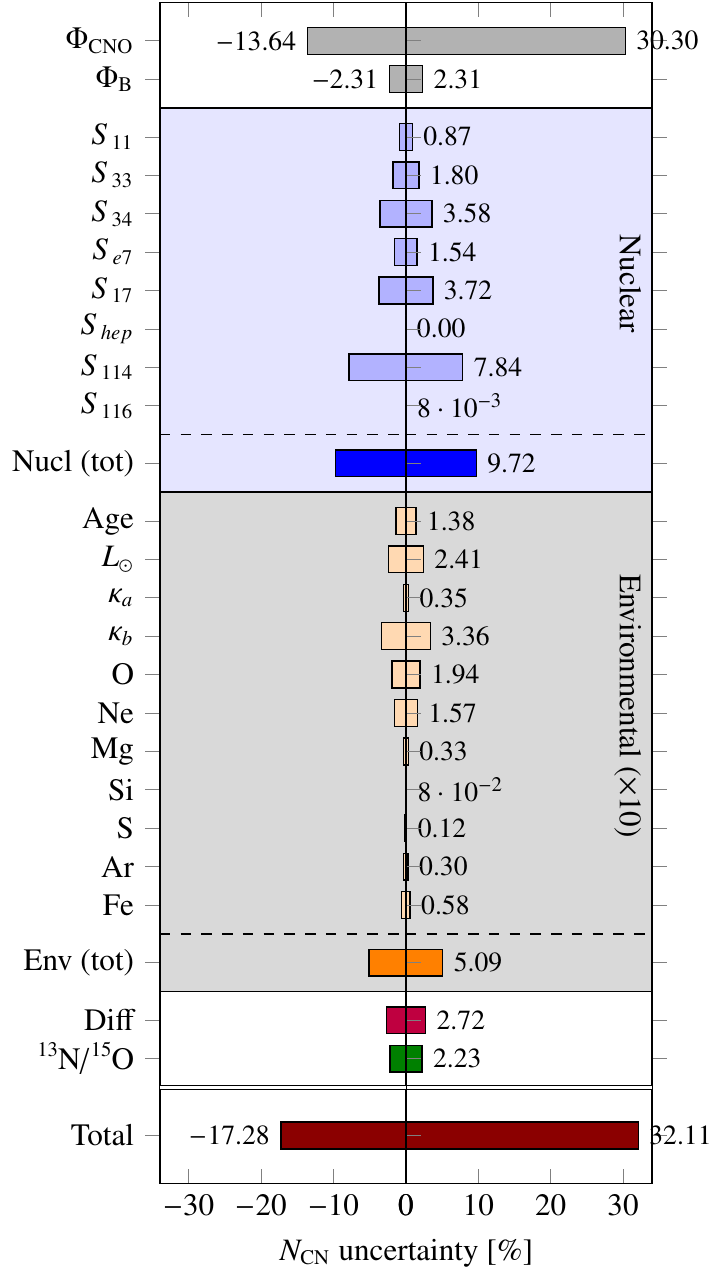}
    \caption{Contributions to the $N_{\ce{CN}}$ error budget. The first two lines indicate the uncertainties linked to the experimental measurement of solar $\nu$ fluxes, the second group to the uncertainty of the nuclear cross-sections, and the third group shows the suppressed contribution of environmental parameters (inflated by $\times 10$). The last lines show the impact of diffusion and of the precision of $\ce{^{13}N}/{\ce{^{15}O}}$ ratio predicted by SSM, as well as the total uncertainty. 
    }
    \label{fig:sm_breakdown}
\end{figure}

The full breakdown of the error budget is shown in Fig.\,\ref{fig:sm_breakdown}.
We notice that the precision of our estimate is limited primarily by the 
$\Phi_{\ce{CNO}}$ determination accuracy, which is worse than the one of $\FluxB$ 
(contributing for $2.3\%$). The second contribution by relevance 
is the one due to the limited precision of the nuclear cross section, which 
account for a $9.7\%$ uncertainty. The main term of nuclear error budget
comes from $S_{114}$ ($7.8\%)$, which is the slowest reaction of the CNO-cyle and therefore
the one determining its pace. The cross section for \ce{^{8}B} production $S_{17}$ also gives 
a non-negligible contribution of $3.7\%$. 
As expected, the uncertainty of environmental parameters does not affect our result, 
accounting for a marginal 0.5\%, while the uncertainty in the ratio between 
\ce{^{13}N} and \ce{^{15}O} events gives a 2.2\% contribution. Finally, the uncertainty in the 
diffusion parameter accounts for a $2.7\%$ uncertainty in the total error budget.

\end{document}